\documentclass[runningheads]{llncs}
\usepackage[T1]{fontenc}
\usepackage{graphicx}
\usepackage{xcolor}
\usepackage{lipsum}
\usepackage{amsmath,amsfonts}
\usepackage{hyperref}
\usepackage{float}
\usepackage{booktabs}

\newcommand{\modelacro}{{QID$^2$}}
\newcommand{\modelacroNSP}{{QID$^2$}}

\begin{document}
\title{\modelacroNSP: An Image-Conditioned Diffusion Model for $Q$-space Up-sampling of DWI Data}
\titlerunning{Diffusion models for Q-space Up-sampling of DWI}
%
\author{Zijian Chen\thanks{Equal Contribution}, Jueqi Wang$^\star$ and Archana Venkataraman}
\authorrunning{Z. Chen et al.}
%
\institute{Department of Electrical and Computer Engineering, Boston University \\ \email{\{zijianc,jueqiw,archanav\}@bu.edu}}
\maketitle              
\begin{abstract}
We propose an image-conditioned diffusion model to estimate high angular resolution diffusion weighted imaging (DWI) from a low angular resolution acquisition. Our model, which we call \modelacroNSP, takes as input a set of low angular resolution DWI data and uses this information to estimate the DWI data associated with a target gradient direction. We leverage a U-Net architecture with cross-attention to preserve the positional information of the reference images, further guiding the target image generation. We train and evaluate \modelacro on single-shell DWI samples curated from the Human Connectome Project (HCP) dataset. Specifically, we sub-sample the HCP gradient directions to produce low angular resolution DWI data and train \modelacro to reconstruct the missing high angular resolution samples. We compare \modelacro with two state-of-the-art GAN models. Our results demonstrate that \modelacro not only achieves higher-quality generated images, but it consistently outperforms the GAN models in downstream tensor estimation across multiple metrics. Taken together, this study highlights the potential of diffusion models, and \modelacro in particular, for q-space up-sampling, thus offering a promising toolkit for clinical and research applications.

\end{abstract}
\section{Introduction}

Diffusion weighted imaging (DWI) is a non-invasive technique that capitalizes on the directional diffusivity of water to probe the tissue microstructure of the brain~\cite{bernsen2013computed}. A typical DWI acquisition applies multiple magnetic gradients, with the field strength controlled by the b-value and the gradient directions given by the b-vectors. Mathematically, these gradients can be represented by a set of coordinates on the sphere, where the magnitude and direction of each coordinate is related to the corresponding b-value and b-vector, respectively. The domain of all such coordinates is called the q-space~\cite{yeh2021tractography}. In general, a denser sampling of directions in the q-space, also known as the angular resolution, leads to higher quality DWI. For example, higher angular resolution acquisitions can improve the tensor estimation~\cite{michailovich2010fast} and facilitates the progression from single-tensor models~\cite{basser1994mr} to constrained spherical deconvolution models~\cite{tournier2007robust} that estimate an orientation distribution function, which captures more complex fiber configurations. However, increasing the angular resolution also prolongs the acquisition time, which can be impractical in clinical settings. Not only are longer acquisitions more expensive, but they are also difficult for some patients to tolerate, which in turn increases the risk of artifacts due to subject motion~\cite{koh2007diffusion}. Given these challenges, it is necessary to explore computational methods that can achieve high-quality DWI with a minimal number of initial scan directions.

Several studies have applied generative deep learning to DWI data. For example, the work of~\cite{zhao2023better} uses a spherical U-Net to directly estimate the ODF using DWI acquired with only 60 gradient directions. More recently, generative adversarial networks (GANs) have also been used to estimate DWI volumes. Specifically, the work of~\cite{ren2021q} generates DWI for a user-specified gradient direction based on a combination of T1 and T2 images~\cite{ren2021q}. Similarly, the authors of~\cite{tatekawa2024deep} use the Pix2Pix model introduced by~\cite{isola2017image} to synthesize DWI with 6 gradient directions from data originally captured with only 3 gradient directions~\cite{tatekawa2024deep}. Further variants of the GAN, such as CycleGAN and DC$^2$Anet, have been applied to simulate a high b-value image from a low b-value one~\cite{rezaeijo2022feasibility}. Autoencoders have also been used to adjust the apparent b-value~\cite{jha2022single}. While these works are seminal contributions to the field, none of them consider the clinically relevant problem of up-sampling a low angular resolution DWI acquisition.

Diffusion models (DMs) have emerged as powerful tool for image generation. At a high level, DMs work by successively adding Gaussian noise to the input and then learning to reverse this noising process~\cite{ho2020denoising}. DMs have been employed in several medical imaging tasks~\cite{kazerouni2023diffusion}, including image translation between modalities~\cite{lyu2022conversion,meng2022novel,li2023zero}, super-resolution and artifact removal~\cite{chung2022score,wang2023inversesr,xiang2023ddm}, registration~\cite{kim2022diffusemorph}, and segmentation~\cite{rahman2023ambiguous,bieder2023memory}. We will leverage DMs to up-sample the DWI gradient directions, as task which to our knowledge, has not been explored in prior work. 

In this paper, we propose an image-conditioned DM, which we call \modelacroNSP, that can estimate high angular resolution DWI from a low angular resolution acquisition. One highlight is that \modelacro automatically identifies several closest available directions and uses the corresponding images as prior knowledge for generating images from any target direction not included in the initial scan. This target image generation process, carried out using a U-Net based structure conditioned on this prior information, can be seen as an extrapolation based on the identified directions and images. By focusing on the most relevant data, \modelacro solicits more targeted prior information and is more computationally efficient. We train and evaluate \modelacro on DWI curated from the Human Connectome Project (HCP) dataset~\cite{van2013wu}. Our model demonstrates superior performance over GAN-based approaches, particularly when the available low angular resolution images are sparsely distributed across the sphere.

\section{Methods}

\begin{figure}[t]
    \centering
    \includegraphics[width=0.95\textwidth]{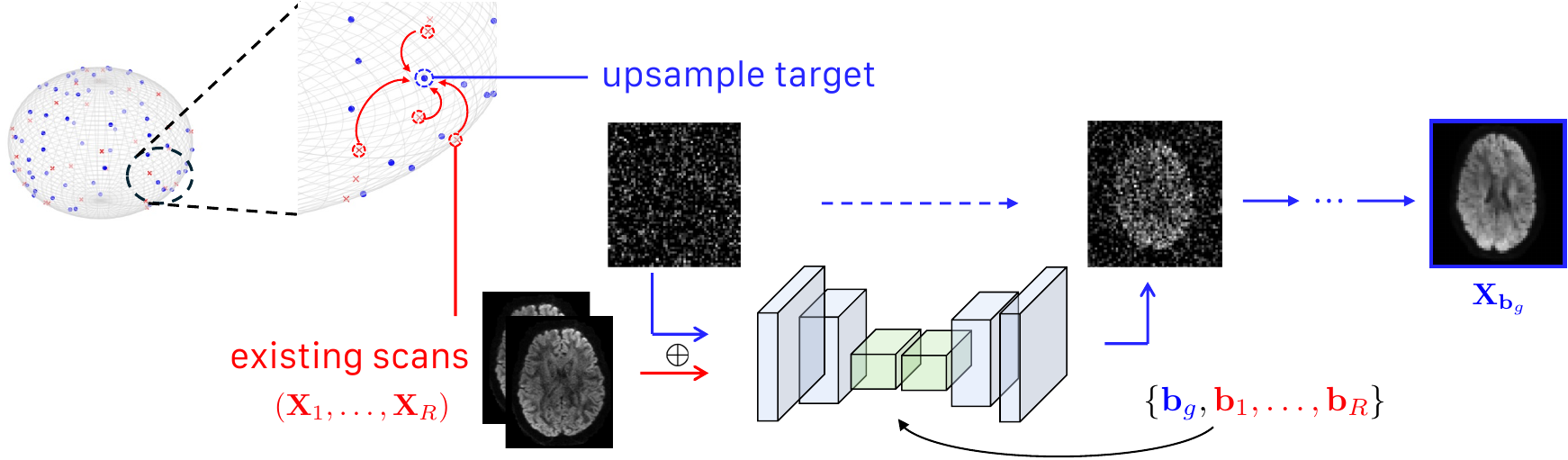}
    \caption{\modelacro framework for up-sampling the angular resolution of DWI. The gray sphere represents the q-space. Red marks are the directions in the low angular resolution scan, and blue marks are the target gradient directions for image generation.}
    \label{fig:model_overview_2}
\end{figure}

Fig.~\ref{fig:model_overview_2} provides an overview of our \modelacro framework. For any user-specified target gradient $\mathbf{b}_g$, our model will find and take as input the $R$ closest reference b-vectors $\mathbf{\bar{b}} = (\mathbf{b}_1, \ldots, \mathbf{b}_R)$ available in the low angular resolution scan and the corresponding DWI slices $\mathbf{\bar{X}} = (\mathbf{X}_1, \ldots, \mathbf{X}_R)$. \modelacro will then output the estimated target image $\mathbf{X}_{\mathbf{b}_g}$. We can obtain a high angular resolution DWI by sweeping the target gradient directions across the sphere and aggregating the generated images with the original low angular resolution scan. 

\subsection{A Diffusion Model for Q-space Up-sampling of DWI}

Inspired by recently-introduced image-conditioned Denoising Diffusion Probabilistic Models (DDPMs)~\cite{wolleb2022diffusion,waibel2023diffusion,durrer2023diffusion}, we design a position-aware diffusion model that leverages ``neighboring" DWI data to estimate the image associated with a target gradient direction. Similar to traditional diffusion models~\cite{ho2020denoising}, \modelacro is comprised of both a forward noising process and a reverse denoising process. 

In the forward process, Gaussian noises are added successively at each time step $t\in \{0,1,\ldots, T\}$ to the generated image. This corruption process is
\begin{equation} \label{equ:forward_process}
    q \left(\mathbf{X}_{\mathbf{b}_g}^{(t)} \mid \mathbf{X}_{\mathbf{b}_g}^{(t-1)}\right) = \mathcal{N}\left(\mathbf{X}_{\mathbf{b}_g}^{(t)} \,; \, \sqrt{1-\beta_t} \mathbf{X}_{\mathbf{b}_g}^{(t-1)} \, , \, \beta_t\mathbf{I} \right), \quad t\geq 1,
\end{equation}
where $\{\beta_t\}$ are the forward process variances, and $\mathbf{X}_{\mathbf{b}_g}^{(t)}$ is the noisy image at time $t$. By repeatedly applying Eq.~\eqref{equ:forward_process} to the starting image $\mathbf{X}_{\mathbf{b}_g}^{(0)} $, we have

\begin{equation}  \label{equ:forward_agg}
    q \left(\mathbf{X}_{\mathbf{b}_g}^{(t)} \mid \mathbf{X}_{\mathbf{b}_g}^{(0)}\right) = \mathcal{N}\left(\mathbf{X}_{\mathbf{b}_g}^{(t)} \, \; \, \sqrt{\overline{\alpha}_t} \mathbf{X}_{\mathbf{b}_g}^{(0)} \, , \, (1-\overline{\alpha}_t)\mathbf{I} \right)
\end{equation}
where ${\alpha}_t=1-\beta_t$ and $\overline{\alpha}_t=\prod_{s=1}^t \, \alpha_s$. Therefore, at step $t$, the generated image $\mathbf{X}_{\mathbf{b}_g}^{(t)}$ can be represented as a function of the initialization $\mathbf{X}_{\mathbf{b}_g}^{(0)}$:
\begin{equation}  \label{equ:forward_x}
    \mathbf{X}_{\mathbf{b}_g}^{(t)} = \sqrt{\overline{\alpha}_t} \, \mathbf{X}_{\mathbf{b}_g}^{(0)} + \sqrt{1-\overline{\alpha}_t} \, \epsilon,\qquad \epsilon\sim \mathcal{N}(0,\mathbf{I}).
\end{equation}

While the forward noising process operates solely on $\mathbf{X}_{\mathbf{b}_g}^{(0)}$, the reference DWI slices~$\{\mathbf{X}_1, \ldots, \mathbf{X}_R \}$ will be used to guide the subsequent denoising process. Rather than constructing a separate network to encode the reference images, which greatly increases the number of parameters and may introduce information loss, we opt to simply concatenate these slices with the target image being generated (i.e., denoised) at each time~$t$ as $\mathbf{\bar{X}}_{\mathbf{b}_g}^{(t)} = \mathbf{X}_{\mathbf{b}_g}^{(t)} \bigoplus_{i=1}^R \mathbf{X}_{\mathbf{b}_i}$.

Starting from the fully corrupted image $\mathbf{\bar{X}}_{\mathbf{b}_g}^{(T)}$, the reverse process aims to gradually recover the original image $\mathbf{\bar{X}}_{\mathbf{b}_g}^{(0)}$. We denote this process as $p_{\theta}(\cdot)$, where $\theta$ denotes the learnable parameters of the underlying neural network. By restricting the denoising to be Gaussian, the process $p_{\theta}(\cdot)$ can be written:
\begin{equation} \label{eq:reverse}
    \small
    p_{\theta}\left(\mathbf{\bar{X}}_{\mathbf{b}_g}^{(t-1)} \mid \mathbf{\bar{X}}_{\mathbf{b}_g}^{(t)} \, ; \,\{\mathbf{b}_g,\mathbf{\bar{b}}\} \right) = \mathcal{N}\bigg(\mathbf{\bar{X}}_{\mathbf{b}_g}^{(t-1)} \, ; \, \mu_{\theta}\left(\mathbf{\bar{X}}_{\mathbf{b}_g}^{(t)},\{\mathbf{b}_g,\mathbf{\bar{b}}\}\right) \, , \, \sigma^2_t \mathbf{I}\bigg),
\end{equation}
where the variances $\sigma_t^2$ are hyperparameters of the model. We note that the denoising process relies on the references DWI data~$\{\mathbf{X}_1, \ldots, \mathbf{X}_R \}$ and the corresponding gradient directions~$\mathbf{\bar{b}} = \{\mathbf{b}_1,\ldots,\mathbf{b}_R\}$, and the target direction~$\mathbf{b}_g$. This combination of inputs allows \modelacro to be position-aware. 

To reverse the forward noising process, we train \modelacro by minimizing the KL-divergence between $p_{\theta}(\cdot)$ and $q(\cdot)$ at each time step~$t$. As shown in~\cite{ho2020denoising} this loss minimization is equivalent to matching the mean functions, i.e.,
\begin{equation} \label{eqn:loss}
    \mathcal{L} = \mathbb{E}_{t,q} \bigg[ \bigg\| \frac{1}{\sqrt{\alpha_t}}\left( \mathbf{\bar{X}}_{\mathbf{b}_g}^{(t)} - \frac{\beta_t}{\sqrt{1-\bar{\alpha}_t}} {\epsilon}\right) - \mu_{\theta}\left(\mathbf{\bar{X}}_{\mathbf{b}_g}^{(t)},\{\mathbf{b}_g,\mathbf{\bar{b}}\}\right) \bigg\|^2 \bigg] .
\end{equation}

The mean function~$\mu_{\theta}(\cdot)$ is generated with a U-Net architecture~\cite{ronneberger2015u,williams2023a} with the cross-attention mechanism based on the concatenated gradient vectors $\mathbf{b}=\begin{bmatrix}
    \mathbf{b}_g & \mathbf{b}_1 & \cdots & \mathbf{b}_R
\end{bmatrix}$. Specifically, the encoding block is computed as follows:
\begin{equation*}
    \mathbf{H}_1 = \mathrm{FF}(\mathbf{H}_0)+\mathbf{H}_0, \quad \mathbf{H}_2 = \mathrm{Attn}(\mathbf{H}_1, \mathbf{b}) + \mathbf{H}_1,    
\end{equation*}
where $\mathrm{FF}(\cdot)$ denotes a feed-forward network, $\mathbf{H}_0$ denotes the block input, and 
\begin{equation*}
    \mathrm{Attn}(\mathbf{H}_1, \mathbf{b}) = \mathrm{Softmax}\bigg(\frac{(W_Q \mathbf{H}_1)(W_K \mathbf{b})^{\top}}{\sqrt{d_k}}\bigg)W_V\mathbf{b},    
\end{equation*}
with $W_Q, W_K, W_V$ being the learned weights and $d_k$ being the dimension of $\mathbf{b}$. The decoding block follows a similar expression but includes skip connections from the corresponding encoding block. This design ensures that image features are effectively attended to and integrated with positional information.

Once \modelacro is trained, we can generate DWI for arbitrary gradient directions by sampling from the standard normal distribution and applying the reverse process in Eq.~(\ref{eq:reverse}) recursively with the corresponding reference images, namely:
\begin{equation} \label{eqn:testing}
    \mathbf{\bar{X}}_{\mathbf{b}_g}^{(t-1)} = \mu_{\theta}\left(\mathbf{\bar{X}}_{\mathbf{b}_g}^{(t)},\{\mathbf{b}_g,\mathbf{\bar{b}}\}\right) + \sigma_t \, \epsilon, \qquad \epsilon\sim \mathcal{N}(0,\mathbf{I}).    
\end{equation}

\subsection{Baseline Comparison Methods}
We compare \modelacro with two state-of-the-art GAN models. 
The first model is a conditional GAN (\textbf{cGAN}) for image generation proposed by~\cite{isola2017image}. We use the same cross-attention U-Net architecture for the generator as used in \modelacroNSP. We use a PatchGAN discriminator~\cite{isola2017image} and inject the gradient direction information~$\{\mathbf{b}_g,\mathbf{\bar{b}}\}$ into the discriminator with cross-attention mechanism. We train the generator to minimize the minimax GAN objective plus a regularization term that encourages voxel-level similarity of the generated and ground-truth images:
\begin{equation*}
    \small
    G^* = \arg \min_{G} \max_{D} \lambda_G \, \left[\mathbb{E}_{\mathbf{X}}\left[\log D(\mathbf{X},\mathbf{b})\right] + \mathbb{E}_{\mathbf{\widetilde{X}}}\left[\log (1 - D(\mathbf{\widetilde{X}}, \mathbf{b}))\right] \right] + \lambda_V \mathcal{L}_1(G)
\end{equation*}
where $\mathbf{X} = [\mathbf{X}_{\mathbf{b}_g}, \mathbf{X}_1,\ldots,\mathbf{X}_R]$ is the concatenated real sample with $\mathbf{X}_{\mathbf{b}_g}$ drawn from the (high resolution) training data and $\mathbf{\widetilde{X}} = [G(\mathbf{X}_{1:R}, \mathbf{b}), \mathbf{X}_1,\ldots,\mathbf{X}_R]$ represents the synthesized data of generated DWI and real reference slices. Finally, $\lambda_{G}$ and $\lambda_{V}$ balance the adversarial and similarity $L_1$ losses, respectively.

The second model is the $Q$-space conditional GAN (\textbf{qGAN}) proposed by~\cite{ren2021q}. Unlike \modelacro and the cGAN baseline, qGAN incorporates the gradient directions and reference DWI data using a feature-wise linear modulation scheme~\cite{perez2018film}. The qGAN discriminator is also a conditional U-Net~\cite{schonfeld2020u} and combines the gradient directions and reference DWI data via an inner product.

Finally, as a sanity check, we compare the deep learning models to a simple interpolation scheme (\textbf{Interp}), in which we express the target gradient direction as a linear combination of the reference gradients and then use the linear coefficients to interpolate between the reference DWI slices to obtain the target.

\subsection{Implementation Details}

For \modelacroNSP, we use a linear noise schedule of 1000 time steps. The \modelacro U-Net employs $[128, 128, 256]$ channels across three levels with one residual block per level. We use the Adam optimizer with a learning rate of $2.5 \times 10 ^{-5}$, $\beta_1=0.5$ and $\beta_2=0.999$. These hyperparameters are selected based on a relevant study~\cite{rombach2022high} and not fine-tuned. We use the same U-Net architecture for the cGAN generator with the same set of hyperparameters. We fix $\lambda_{G} = 1$ and $\lambda_{V}=100$ in the loss function weights for both GAN methods. The discriminator is updated once for every two updates of the generator during training~\cite{ren2021q}. For qGAN, we use a learning rate of $5 \times 10^{-5}$ with the Adam optimizer. For comparison, we evaluate all models with both $R=3$ and $R=6$ reference DWI data.

To avoid memory issues~\cite{pinaya2022brain}, we train the deep learning models to generate 2D axial slices, which we stack into 3D DWI volumes. Each 2D image has a size of $(145, 174)$. During training, we independently normalize each slice from its original intensity to a range of $[0,1]$. Data augmentation is employed to enhance model training. Specifically, we use rotations by random angles in $[- 15^{\circ}, 15^{\circ}]$ and random spatial scaling factors in $[0.9, 1.1]$. The final output is rescaled voxel-wise back to the original intensity and masked by the subject $\mathbf{X}_{\mathbf{b}_g}$ image.

\section{Experimental Results}

\textbf{Dataset Curation and Preprocessing:}
We curate a total of 720 subjects from the HCP S1200 release~\cite{van2013wu}. The remaining HCP subjects are excluded due to an inconsistent number of gradient directions at $b=1000 \mathrm{~s/mm^2}$. The DWI is acquired on a Siemens 3T Connectome scanner at 3 shells ($b=1000,2000$ and $3000~s/mm^2$). Each shell has exactly 90 gradient directions sampled uniformly on the sphere. The voxel size is $1.25 \times 1.25 \times 1.25\mathrm{~mm}^3$. The data is preprocessed with distortion/motion removal and registration to the 1.25$~\mathrm{mm}$ structural space.

Clinical diffusion imaging typically uses lower b-values with approximately 30 gradient directions~\cite{doshi2017physiologic}. To better accommodate this situation, we focus our evaluation on the $b=1000\mathrm{~s/mm^2}$ shell. From here, we construct low angular resolution DWI data by subsampling the 90 gradient directions to 30 evenly spaced ones that preserve the uniformity of the sphere~\cite{cheng2017single}. The data for the remaining 60 directions serve as the targets for model training and evaluation. 

\begin{figure}[t]
    \centering
    \includegraphics[width=0.9\textwidth]{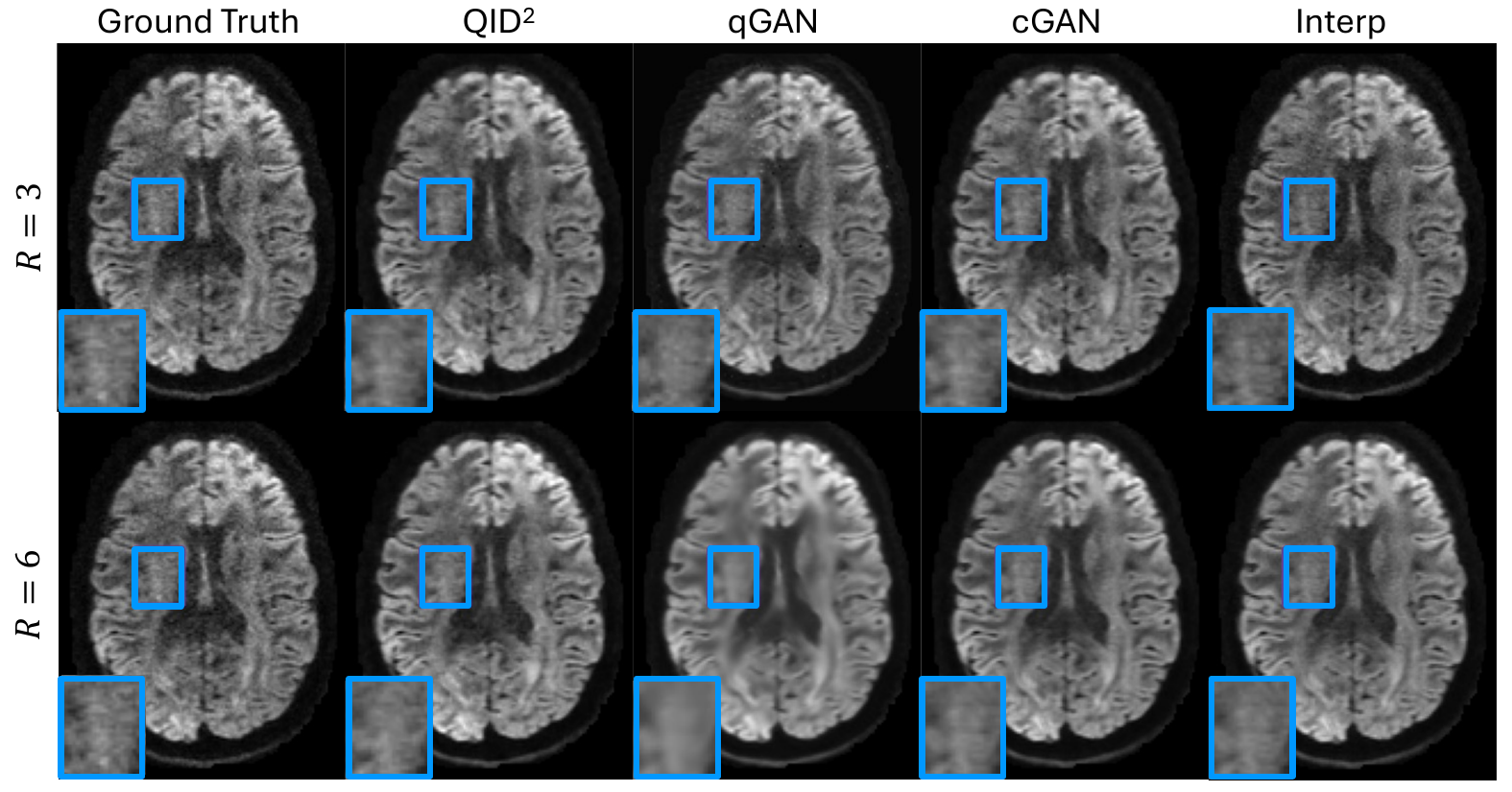}
    \caption{Qualitative results that compare the ground-truth DWI acquisition to images generated by \modelacro and the baselines methods for $R=3$ and $R=6$. Zoomed-in area highlights details that are preserved by our method and do not appear in the baselines.}
    \label{fig:DWI_qual}
\end{figure}

Each volume is broken down into 145 axial slices. The deep learning models are trained to predict the image slices for each target direction. Based on this scheme, we create 60 samples for each slice. Each sample consists of one 2D slice for the target gradient direction and $R$ reference slices corresponding to the closest low resolution gradient directions. The distance between gradients is defined by the geodesic distance on the sphere: $d(\mathbf{b}_1,\mathbf{b}_2) = \arccos(\mathbf{b}_1\mathbf{b}_2^{\top})$.

Finally, we use 576 HCP subjects for training, 72 for validation, and 30 for testing. The original DWI scans are treated as the gold standard for evaluation.

\medskip \noindent
\textbf{Comparing Reconstructed Image Quality:}
Fig.~\ref{fig:DWI_qual} presents qualitative results that compare the ground-truth DWIs to those generated by \modelacro and the baseline methods. The GAN models and Interp fail to preserve high-frequency details in the synthesized DWI data, while \modelacro succeeds in capturing the finer details more accurately, as highlighted in the zoomed-in blue boxes.

Table~\ref{tab:DWImetric} (left) reports the Fréchet inception distance (FID)~\cite{heusel2017gans} and the structural similarity index measure (SSIM)~\cite{1284395} of the synthesized DWI data. We observe that \modelacro achieves nearly a two-fold improvement (i.e., decrease) in FID than the GAN models for both $R=3$ and $R=6$, which indicates that the DWI data generated by diffusion possess higher quality and greater diversity. Although the GAN models achieves a slightly higher SSIM than \modelacroNSP, the difference is not statistically significant using a two-sample (paired) $t$-test. Interestingly, the simple interpolation technique achieves better FID than \modelacro when $R=3$. This is likely because the interpolation tracks the closest reference image, which is more akin to the original DWI distribution.
However, the improved FID does not generalize to better tensor estimation, as seen in the next section.

\begin{table}[t]
    \setlength{\tabcolsep}{5pt}
    \centering
    \caption{Quantitative evaluation of the generated image quality (left) and FA estimation quality (right) of \modelacroNSP, the GAN models, and interpolation for different $R$. The best performance of each metric is highlighted in bold.}
    \label{tab:DWImetric}    
    \begin{tabular}{c|cc|cc}
        \toprule
        Methods & Image FID $\downarrow$ & Image SSIM $\uparrow$ & FA Error $\downarrow$ & FA Map SSIM $\uparrow$ \\
        \midrule
        \modelacro($R$=3) & 14.07 &  0.895 $\pm$ 0.045 & \textbf{0.027 $\pm$ 0.003} & $\textbf{0.866} \pm \textbf{0.043}$ \\
        qGAN($R$=3) & 24.85 &  0.893 $\pm$ 0.046 & $0.037 \pm 0.002 $ & $0.792 \pm 0.052$\\
        cGAN($R$=3) & 29.93 &  0.913 $\pm$ 0.039 & $0.099 \pm 0.014$ & $0.643 \pm 0.159$\\
        Interp($R$=3) & \textbf{8.96} &  \textbf{0.917 $\pm$ 0.038} & $0.057 \pm 0.026$ & $0.750 \pm 0.110$\\
        \hline
        \addlinespace[0.2em]
        \modelacro($R$=6)  & \textbf{16.29} &  0.900 $\pm$ 0.045 & $\textbf{0.027} \pm \textbf{0.003}$ & $\textbf{0.863} \pm \textbf{0.042}$ \\
        qGAN($R$=6)      & 71.44 &  0.905 $\pm$ 0.042 & $0.040 \pm 0.004$ & $0.801 \pm 0.045$\\
        cGAN($R$=6)       & 36.33 &  0.915 $\pm$ 0.037 & $0.031 \pm 0.004$ & $0.851 \pm 0.044$\\
        Interp($R$=6) & {21.46} & \textbf{0.933 $\pm$ 0.044} & $0.038 \pm 0.006$ & $0.815 \pm 0.052$ \\
        \bottomrule
    \end{tabular}
\end{table}

\begin{figure}[t]
    \centering
    \includegraphics[width=0.95\textwidth]{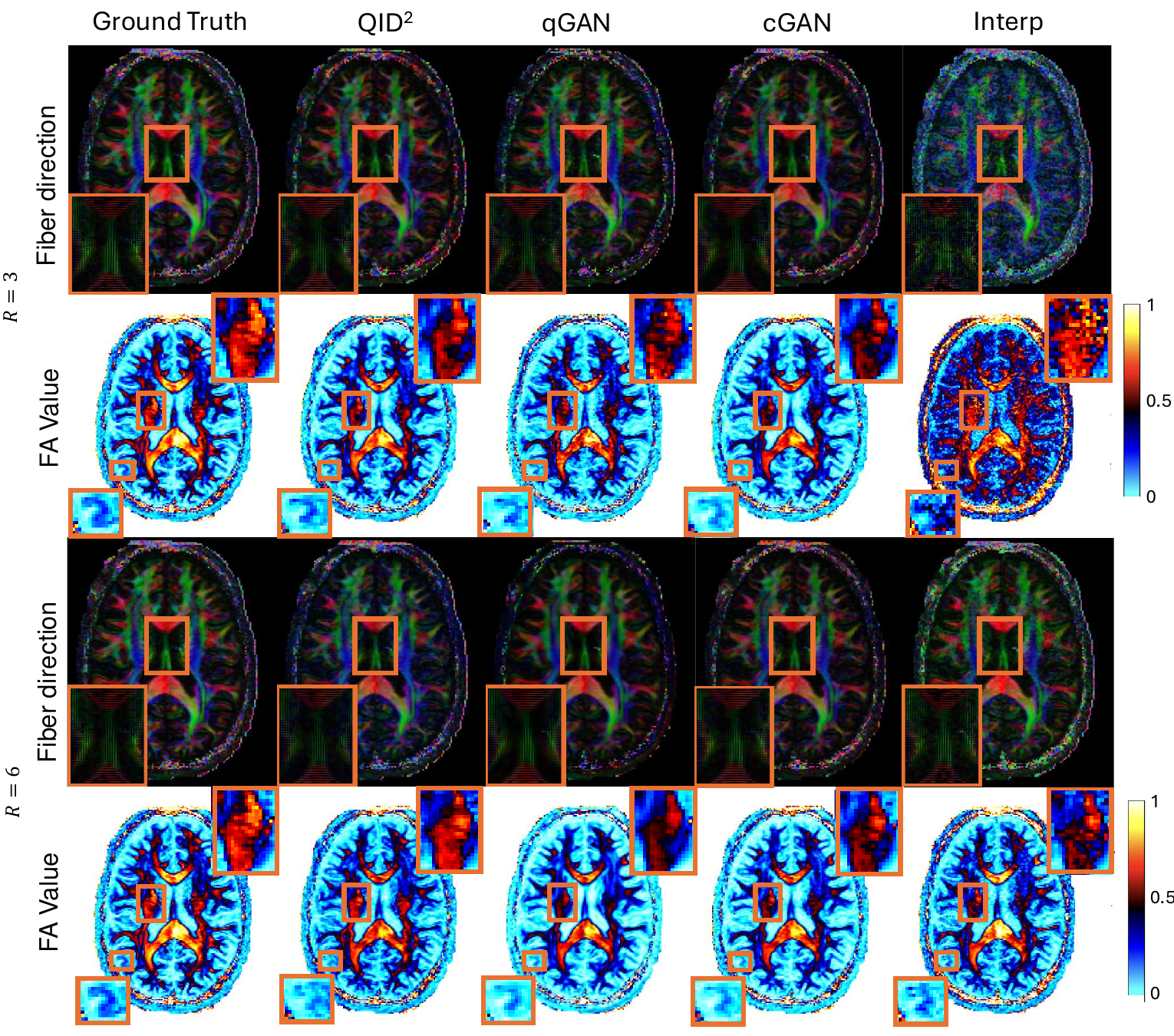}
    \caption{Fiber direction and FA value map estimated based on the ground truth images, \modelacroNSP, and baseline methods. \textbf{Row 1/3:} Colored FA maps indicating fiber orientation, with minimal visually detectable differences among the images.  \textbf{Row 2/4:} FA value maps, where brighter colors indicate higher FA values. Significant differences compared to the ground truth are zoomed-out with orange boxes.}
    \label{fig:FA_plt}
\end{figure}

\medskip \noindent
\textbf{Impact on Tensor Estimation:}
We estimate the fractional anisotropy (FA) using the standard single-tensor model~\cite{basser1994mr}. Fig.~\ref{fig:FA_plt} shows the fiber direction and FA value maps among the ground-truth, \modelacro and baseline methods for $R=3$ and $R=6$. Similarly to the finding in the reconstructed image, we observe that the qGAN and cGAN methods capture the general FA trends but fail to capture the high frequency features. Conversely, the diffusion-generated image by \modelacro more closely resembles the ground-truth data by capturing finer details more accurately. This shows that the visual differences in the reconstructed images in Fig.~\ref{fig:DWI_qual} are important when estimating tensors. The Interp method fails to generate realistic FA maps for $R=3$. Empirically, we also observe quality issues with Interp for $R=6$ even though they are less evident in the figure.

Table~\ref{tab:DWImetric} (right) reports the mean absolute error and SSIM, as compared to the FA computed from the ground-truth high angular resolution DWI. As seen, \modelacro consistently outperforms the GAN-based model and the Interp method for both $R=3$ and $R=6$. Specifically for $R=3$, the error in FA is roughly three times lower for \modelacro than for the GANs. \modelacro also achieves significantly higher SSIM values. These trends persist when the number of reference images increases to $R=6$, i.e., even when more prior information is provided. However, the relative performance gain over the GANs shrink. Additionally, although the image-based metrics are better for the interpolation-generated (Interp) images, \modelacro outperforms this baseline by a large margin when estimating FA. Taken together, these results suggest that \modelacro is particularly effective in scenarios where the images are scarce and distributed sparsely, i.e., smaller values of $R$.

\section{Conclusion}
We introduce an image-conditioned diffusion model (\modelacroNSP) that can generate high angular resolution DWI from low angular resolution data, effectively estimating high-quality imaging with limited initial scan directions. Our approach takes advantage of similar DWI data as prior information to predict the data for any user-specified gradient direction. The results demonstrate that diffusion-generated DWIs by \modelacro achieve superior quality and significantly outperform those generated by GAN models in downstream tensor modeling tasks. Although our method currently exhibits longer training times due to the denoising characteristics of DDPMs, this limitation could be mitigated by employing more efficient sampling techniques~\cite{song2020denoising} in future work. We will also focus on on clinical datasets to evaluate its robustness in real-world clinical scenarios. 

\subsection*{Acknowledgments}

This work was supported by the National Institutes of Health R01 HD108790 (PI Venkataraman), the National Institutes of Health R01 EB029977 (PI Caffo), the National Institutes of Health R21 CA263804 (PI Venkataraman).
%

%
%
%
\bibliographystyle{splncs04}
\bibliography{d2}

\end{document}